\documentclass[%
 aip,
 amsmath,amssymb,
 reprint,%
]{revtex4-1}

\usepackage{graphicx}
\usepackage{dcolumn}
\usepackage{bm}

\usepackage[utf8]{inputenc}
\usepackage[T1]{fontenc}
\usepackage{mathptmx}
\usepackage{etoolbox}
\usepackage{xcolor}

\makeatletter
\def\@email#1#2{%
 \endgroup
 \patchcmd{\titleblock@produce}
  {\frontmatter@RRAPformat}
  {\frontmatter@RRAPformat{\produce@RRAP{*#1\href{mailto:#2}{#2}}}\frontmatter@RRAPformat}
  {}{}
}%
\makeatother
\begin{document}

\preprint{AIP/123-QED}

\title{A Permutation-equivariant Deep Learning Model for Quantum State Characterization}
\author{D. Maragnano}
\email{diego.maragnano01@universitadipavia.it}
\affiliation{%
Dipartimento di Fisica, Università di Pavia, via Bassi 6, 27100 Pavia, Italy
}
\author{C. Cusano}
\affiliation{%
Dipartimento di Ingegneria Industriale e dell'Informazione, Università di Pavia, via Ferrata 5, 27100 Pavia, Italy
}
\author{M. Liscidini}%
\affiliation{%
Dipartimento di Fisica, Università di Pavia, via Bassi 6, 27100 Pavia, Italy
}

\date{\today}

\begin{abstract}
The characterization of quantum states is a fundamental step of any application of quantum technologies. Nowadays there exist several approaches addressing this problem, also based on machine and deep learning techniques. However, all these approaches usually require a number of measurement that scales exponentially with the number of parties composing the system. Threshold quantum state tomography (tQST) addresses this problem and, in some cases of interest, can significantly reduce the number of measurements. In this paper, we study how to combine a permutation-equivariant deep learning model with the tQST protocol. We test the model on quantum state tomography and purity estimation. Finally, we validate the robustness of the model to noise. We show results up to 4 qubits.
\end{abstract}

\maketitle


\section{Introduction}

Quantum state characterization is pivotal in the development and implementation of quantum technologies, from quantum computing to quantum cryptography and sensing\cite{krenn2023artificial}. At its core, quantum state characterization involves determining the representation of a quantum system, through processes such as quantum state tomography (QST), or estimate quantities related to the state, such as the purity, the fidelity with a given state, or the degree of entanglement.

QST aims at reconstructing the representation of a quantum state from the measurement of a large enough number of observables. Many standard approaches solve this task \cite{james_2001, baumgratz2013scalable, blume2010optimal, gupta_maxent, sg_qst}, all suffering from the same drawback, sometimes called the curse of dimensionality of QST: the number of observables required for the reconstruction of a quantum state of \emph{n} qu\emph{d}its, \textit{i.e.}, \emph{n} \emph{d}-level carriers of information, scales exponentially with the number of parties as $d^{2n}$. In this respect, some methods have been developed to tackle this problem, such as the compressed sensing and the permutationally invariant one\cite{gross_cs, adaptive_noapriori, adaptive_numerical, toth2010permutationally, schwemmer2014experimental}.

Threshold quantum state tomography (tQST) is a recently developed approach that can address and solve this problem in an efficient way without any assumption on the state\cite{binosi2024tailor}. Furthermore, it is deterministic, in the sense that the user knows how many resources are needed to conclude the protocol, in contrast to adaptive methods\cite{mahler2013adaptive, quek2021adaptive, lange2023adaptive}.
Shadow tomography in turn focuses on estimating properties that are function of the state without reconstructing the entire state\cite{aaronson2018shadow, huang2020predicting, huang2022provably, acharya2021shadow, wei2024neural, kokaew2024bootstrappingclassicalshadowsneural}.

Machine and deep learning-based techniques have been used for quantum state characterization too. They range from traditional deep learning models, such as feed-forward \cite{koutny2022neural, xin2019local, gao2018experimental, palmieri2020experimental, cai2018approximating}, convolutional \cite{schmale2022efficient, lohani2020machine}, and recurrent neural networks\cite{quek2021adaptive}, 
\textcolor{black}{to physics-inspired ones, such as restricted Boltzmann machines\cite{torlai2018neural, neugebauer2020neural, PhysRevLett.123.230504, nn_states_string, carleo2018constructing}. Models within the generative framework \cite{ahmed2021quantum, zhu2022flexible, carrasquilla2019reconstructing, li2024experimental, PhysRevResearch.3.033278, rocchetto2018learning} and attention-based ones\cite{palmieri2024enhancing, cha2021attention} have been considered too.}

\textcolor{black}{In this work we study if it is possible to combine the tQST protocol with deep learning models to carry out two quantum state characterization tasks in realistic scenarios}. First, we develop a permutation-equivariant model, and train it to reconstruct density matrices according to the tQST protocol.
Then, we train the same model to estimate the purity of quantum states. Finally, we validate the robustness of our model against depolarizing and state-error noise, and show results for 2 and 4 qubits.

The paper is organized as follows. In Section \ref{sec:tqst_protocol} we illustrate the tQST protocol, and how to include neural networks into it. In Section \ref{sec:models} we provide details about the permutation-equivariant model, the datasets. In Section \ref{sec:results} we show numerical results for 2 and 4 qubits. Finally, we draw our conclusions.

\section{Threshold Quantum State Tomography Protocol}\label{sec:tqst_protocol}

Conventional QST aims at reconstructing the representation of a quantum state from the measurement of a sufficiently large number of observables. The representation of quantum states considered in this work is the density matrix, \textit{i.e.}, a trace-one, Hermitian, and positive semi-definite matrix. The number of observables needed for this task is dictated by the dimension of the density matrix. In particular, a system of \( n \) qubits requires at least \( 4^n -1 \) observables, equal to the number of real independent parameters characterizing the density matrix. One can perform the measurements in any sequence, and then process the acquired information via proper statistical inference methods, such as maximum likelihood or Bayesian mean estimation\cite{james_2001, blume2010optimal}.

The three properties of density matrices, in particular the positive semi-definiteness, imply that any of its off-diagonal elements \( \rho_{ij} \) must satisfy the relation \( |\rho_{ij}| \leq \sqrt{\rho_{ii} \rho_{jj}} \). The measurement of the diagonal elements of the density matrix then provides information about the off-diagonal ones. For instance, if \( \rho_{ii} \) is vanishing, then all elements in the \( i \)-th row and column of \( \rho \)  are vanishing too. Similarly, if \( \rho_{ii} \) and \( \rho_{jj} \) are non-vanishing but small with respect to the other diagonal elements, then the modulus of \( \rho_{ij} \) will also be small.

These observations are the core of tQST, the protocol of which we now illustrate\cite{binosi2024tailor}. First, one measures the \textcolor{black}{$2^n$} diagonal elements \( \{\rho_{ii}\} \) of the density matrix. One can directly measure \textcolor{black}{these} elements by projecting onto the \textcolor{black}{$2^n$} elements of the chosen computational basis. \textcolor{black}{QST methods that do not rely on assumptions on the state require this measurement, even if it is exponential in the number of qubits.} Second, a threshold \( t \) is chosen, and using the information from \( \{\rho_{ii}\} \), one identifies those \( \rho_{ij} \) for which \( \sqrt{\rho_{ii} \rho_{jj}} \geq t \). Third, a proper set of observables associated with these \( \rho_{ij} \) is constructed, and one performs only those measurements. Finally, one process the results via a statistical inference tool to reconstruct the density matrix.

There are some important points to highlight about the tQST protocol.
\begin{enumerate}
    \item  The tQST protocol is deterministic. The resources to complete the experiment are determined once the threshold is chosen. On the contrary, adaptive approaches do not display this feature, for each measurement is chosen based on the previous one\cite{adaptive_noapriori, adaptive_numerical}.
    \item The tQST approach does not make any a priori assumption about the state to be reconstructed. Other methods can reduce the number of measurements, or improve the scaling with the dimension of the system, by making hypothesis on the structure of the state\cite{gross_cs, baumgratz2013scalable}.
    \item The main purpose of the threshold \( t \) is to control the resources required by the protocol, such as the time to perform the experimental measurements or the computational power to store and process the data. By setting \(t>0\), one may need fewer resources than those required by conventional QST.
    \item In principle, the threshold \( t \) can be set by the user according to the available resources, as discussed in the previous point.
    However, the tQST protocol suggests that the reduction in the number of measurements should be particularly relevant for sparse matrices. Thus we related the threshold \(t \) to the sparsity of the density matrix by deriving a closed formula for \(t\) as a function of the initial measurements required by the protocol, \textit{i.e.}, the diagonal elements of the density matrix. To this end, we first adopted the Gini index\textcolor{black}{\cite{5238742}.} 
    Let $\underline{c} = \left[ c_{(1)}, c_{(2)}, \cdots, c_{(N)} \right]$, such that $c_{(1)} \le c_{(2)} \le \cdots \le c_{(N)}$ and \(c_{(i)} \geq 0 \hspace{.1cm} \forall i\). The Gini index is defined as:
    \begin{align}
        \text{GI} \left( \underline{c} \right) = 1 - 2 \sum_{k=1}^N \frac{c_{(k)}}{\| \underline{c} \|_1} \left( \frac{N-k+\frac{1}{2}}{N} \right),
    \end{align}
    with \(\| \underline{c} \|_1 = \sum_i c_i\). It satisfies \( 0 \le \text{GI} \left( \underline{c} \right) \le 1 - 1/N \).
    We adapted this definition to make the Gini index a suitable threshold for tQST. Based on the tQST protocol, the vector to compute the Gini index is the diagonal of the density matrix. Thus $N = 2^n$ and the threshold is set to:
    \begin{align}
        t = \frac{\text{GI} \left( \underline{\rho} \right)}{2^n-1},
    \end{align}
    with $\underline{\rho} = \left( \rho_{11}, \rho_{22}, \cdots, \rho_{NN} \right)$. We do not know yet if this value of the threshold is optimal, but our results suggest it could be. We will investigate this aspect in future studies.
\end{enumerate}

While in the original work of tQST\cite{binosi2024tailor} the authors adopted maximum likelihood estimation for the reconstruction of the density matrix, the purpose of this work is different. On the one hand, we develop and use a physics-inspired deep learning model to retrieve the density matrix from the measurements provided by the tQST protocol. \textcolor{black}{Our aim is to explore the feasibility of deep-learning-based data-processing as an alternative to maximum likelihood, without reducing the number of measurements.} On the other hand, we use the same model and measurements to estimate the purity of quantum states, thus extending the applicability of the tQST protocol to other tasks than only density matrix reconstruction.

\section{Models \& Datasets}\label{sec:models}

\subsection{Deep Learning Models}

The first model we tested for our purposes was a Multi-Layer Perceptron (MLP), as the simplest deep learning architecture to implement, and due to its behavior as universal approximator and high flexibility. \textcolor{black}{It comprises a sequence of Linear layers (according to Pytorch nomenclature), each followed by an activation function.}

\textcolor{black}{We then developed a permutation-equivariant MLP (PEMLP) and applied it to threshold quantum state tomography and purity estimation.} The motivation behind this choice is the following. Let us consider a density matrix \(\rho\) and an observable \(O\) represented in a given basis \(\mathfrak{B}\). The possible outcomes of a measurement of \(O\) are the corresponding eigenvalues, and the expectation value of \(O\) is given by \( \langle O \rangle = \text{Tr} \left[ \rho O \right]\). Let us now permute, \textit{i.e.,} reorder, the elements of \(\mathfrak{B}\) to get a permuted basis \(\mathfrak{B}'\), and accordingly find the permuted representations of the density matrix \(\rho'\) and the observable \(O'\). A permutation of the basis does not change the eigenvalues of \(O\) and \(\rho O\), thus the physical results as the outcomes of a measurement or the expectation value of the observable are unchanged, as expected. 

This feature of density matrices allowed us to introduce permutation-equivariant \textcolor{black}{linear (PELinear)} layers as part of the architecture of our \textcolor{black}{PEMLP} models. In particular, we used them to infer the density matrix or the purity of the state from the measurement outcomes. Details on the implementation \textcolor{black}{of the PELinear layers} and the training can be found in Appendix \ref{app:nns} and \ref{app:training_details}, respectively.

\subsection{Noiseless Datasets}\label{subsec:noiseless_datasets}

We generated separated datasets for the density matrices, the corresponding purities, and the measurement outcomes according to the tQST protocol.

The first dataset we constructed was the one containing the density matrices, which are the object to reconstruct in quantum state tomography. The $n$-qubit density matrix dataset was generated such that it contains the same share of pure and mixed states. In the case of mixed density matrices, we first randomly choose the number of vanishing diagonal elements $z$ between 0 and $2^n -2$. This determines the possible values of the rank $r$ of the mixed density matrix, for it varies between 2 and $2^n - z$. The number of possible pairs $\left( z, r \right)$ is $ \sum_{k=1}^{2^n -1} k = 2^{n-1} \left( 2^n -1 \right)$. For each pair $\left( z, r \right)$ we generated $M$ random density matrix. This the number of mixed density matrices is $M \times 2^{n-1} \left( 2^n -1 \right)$. Finally, we generated an equal number of pure density matrices fixing the rank $r = 1$ and choosing $z$ as before. The total number of density matrices in the dataset is then $2 \times M \times 2^{n-1} \left( 2^n -1 \right)$. The 2-qubit dataset consisted of 24,000 density matrices $\left( n=2, M=2000 \right)$, while the 4-qubit dataset of 120,000 density matrices $\left( n=4, M=500 \right)$.

We then constructed the dataset with the measurement outcomes, which will be the inputs of our models. For each density matrix we computed the measurement outcomes according to the tQST protocol\cite{binosi2024tailor}. If $\sqrt{\rho_{ii} \rho_{jj}} \ge t$ the measurement is performed and the outcome is recorded, otherwise the measurement is not performed and we record a mock value 2. We choose to use 2 to represent the non-performed measurements for it is out of the range of the valid measurement outcomes, which are positive real numbers between 0 and 1. Finally, for each density matrix $\rho$ we computed and stored the corresponding purity as $\mathcal{P} = \text{Tr} \left[ \rho^2 \right]$, thus constructing the datasets for the purity estimation.

\subsection{Noisy Datasets}\label{subsec:noisy_datasets}

We generated noisy datasets to study the impact of noise on the performance of our models. In particular, we studied if the models could reconstruct the noiseless density matrix, or estimate the corresponding purity, from noisy measurement outcomes. To generate the required datasets, we applied a noisy channel to the noiseless density matrix, and then computed the noisy measurement outcomes. We remark that the number of performed measurements is in general different from the noiseless case, for the diagonal elements of the density matrix change after the application of the noisy channel.

We considered two noisy channels. The first is the depolarizing channel, which maps a density matrix $\rho$ into a linear combination of itself and the maximally entangled state:
\begin{align}
    \mathcal{E}_{\text{depol}} \left( \rho \right) = \left( 1-p \right) \rho + \frac{p}{2^n} \mathbb{I}_{2^n \times 2^n}.
\end{align}
with $0 \le p \le 1 + 1/(d^2 -1)$ a parameter quantifying the strength of the noise, and $\mathbb{I}_{2^n \times 2^n}$ the $2^n$-dimensional identity matrix.
For a single qubit, the depolarizing channel has a clear geometric meaning: it represents a uniform contraction of the Bloch sphere. The second channel was the experimental-state error\cite{PhysRevA.79.022109}, which maps a density matrix $\rho$ into a convex combination of itself and a random density matrix:
\begin{align}
\begin{split}
    \mathcal{E}_{\text{exp-state}} \left( \rho \right) &= \left( 1 - \varepsilon \right) \rho + \varepsilon \rho_{\text{random}} \\
    \rho_{\text{random}} &= \frac{R^{\dagger} R}{\text{Tr} \left[ R^{\dagger} R \right]} \\
    R &= 2 \text{rand} \left( 2^n \right) -1 +i \left[ 2 \text{rand} \left( 2^n \right) -1 \right].
\end{split}
\end{align}
The $\text{rand} \left( 2^n \right)$ function generates a $2^n \times 2^n$ matrix of random values sampled from a uniform distribution over the interval (0,1). This channel simulates experimental error in the preparation of the state. Indeed, the prepared state always differs from the intended state by some small amount, here quantified by the $\varepsilon$ parameter and implemented via a random density matrix. 

\subsection{Workflow of AI-based tQST}

The workflow of AI-based tQST proceeds as follows: we generate a (noiselesse or noisy) density matrix and perform the measurement specified by the tQST protocol. The measurement outcomes are recorded as explained in \ref{subsec:noiseless_datasets}, stored in a 1d array of size $\left( 1, 4^n \right)$, and fed into the model for the inference task.

The architectures of the models are similar for both tasks. They consists of an input layer of $4^n$ neurons, a number of hidden layers (which depends on the size of the density matrix), and a final output layer with $4^n$ neurons for tQST, while a single neuron for the purity estimation. 

The output of the model has a different size depending on the performed task. For tQST, the output array contains the real parameters characterizing the density matrix, including first the (real) diagonal elements, then the off-diagonal elements (real and imaginary parts interleaved). The one-dimensional output array is then converted into a Hermitian, trace-one matrix $\mu$. However, this reconstruction process does not guarantee $\mu$ to be positive semi-definite. We handle this problem by numerically implementing the solution outlined in \textcolor{black}{\cite{maxlik_mineff}}. For the purity estimation, the output array contains only one element. The model can estimate a purity value slightly larger than 1 or smaller than the minimum $1/d$, with $d$ the dimension of the Hilbert space associated to the system. In this case, the estimated values are rescaled to the maximum or minimum, respectively.

\section{Results \& Discussion}\label{sec:results}

\subsection{2 Qubits}

In this section we show the results of our method for the two-qubits case. This is considered as a benchmark for the validation of the method. We tested the \textcolor{black}{MLP} and \textcolor{black}{PEMLP} models for threshold quantum state tomography and to estimate the state purity.

For the tomography task, the \textcolor{black}{MLP} consists of two hidden \textcolor{black}{Linear} layers, each containing 32 neurons, while the \textcolor{black}{PEMLP} has a single hidden \textcolor{black}{PELinear} layer with 64 internal features. In both models, the activation function between hidden layers is the ReLU function \cite{hahnloser2000digital, glorot2011deep}. The performance on the test set was evaluated using the average fidelity as a figure of merit, defined as:
\begin{align}
F = \frac{1}{N_{test}} \sum_i \text{Tr} \left( \sqrt{ \sqrt{\rho_{o,i}} \rho_{p,i} \sqrt{\rho_{o,i}}  } \right),
\end{align}
with $\rho_o$ $\left( \rho_p \right)$ the observed (predicted) density matrix, and $N_{test}$ the number of samples in the test set. The \textcolor{black}{MLP} achieved an average fidelity of $F = 0.9555 \pm 0.0411$ on the noiseless dataset, while the \textcolor{black}{PEMLP} achieved $F = 0.8893 \pm 0.0919$. These results demonstrate that both models can reconstruct the density matrix of two qubits from measurement outcomes collected according to the tQST protocol. 
The results are compatible with those obtained with maximum likelihood, which achieved an average fidelity of $F= 0.9941 \pm 0.0174$ on the test set.

We then investigated the impact of noise on the accuracy with which the state can be reconstructed to assess the practical applicability of these models in an experimental scenario, as explained in \ref{subsec:noisy_datasets}. The strengths $p$ and $\varepsilon$ of the noisy channels were set to 0.01, 0.05, and 0.1 in our experiments. The results, reported in Appendix \ref{app:2qubit_results}, indicate that both the \textcolor{black}{MLP} and the \textcolor{black}{PEMLP} are robust against noise, with slightly modified performance as noise strength increases.

We further analyzed the models' performance based on the number of zeros on the diagonal of the noiseless density matrix. For each model, we computed the average fidelity for density matrices with 0, 1, and 2 vanishing diagonal elements separately. The \textcolor{black}{MLP} performed consistently well across these subsets, whereas the \textcolor{black}{PEMLP} showed slightly better performance on density matrices with two vanishing diagonal elements. These findings suggest that both models are reliable in reconstructing sparse states, such as Bell states.

\begin{figure*}
    \includegraphics[scale=0.3]{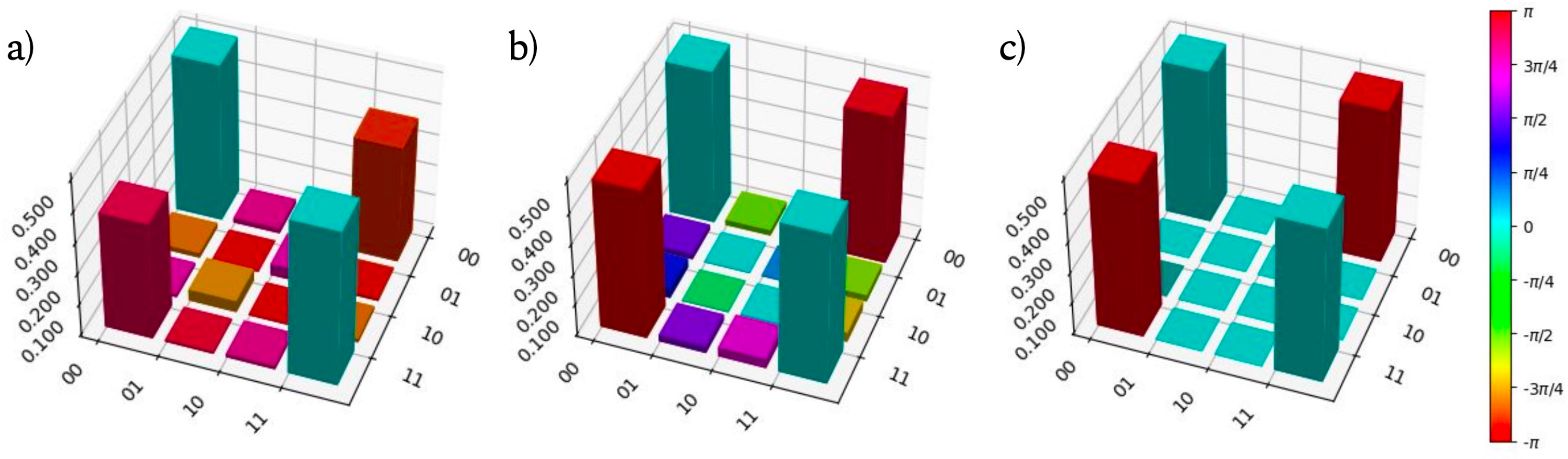}
    \caption{Reconstruction of the density matrix of the Bell state $\lvert \phi^- \rangle$. The height of the bar represents the absolute value, while the phase is encoded in the color as shown in the colorbar. Measurement outcomes are experimental data from Ref. \textcolor{black}{\cite{altepeter20044}}. a) tQST reconstruction with the \textcolor{black}{MLP} model trained on the depolarizing noise dataset with $p=0.01$. The fidelity with the noiseless target state is 0.93. b) tQST reconstruction using maximum likelihood. The fidelity with the noiseless target state is 0.99. c) Target density matrix.}
    \label{fig:dm_example}
\end{figure*}

We substantiate this by reconstructing the Bell state $\lvert \phi^- \rangle = 1/\sqrt{2} ( \lvert 00 \rangle - \lvert 11 \rangle )$ using high-quality experimental data from \textcolor{black}{\cite{altepeter20044}}, where the authors perform the tomography of single- and two-qubit states encoded in the polarization of photons generated via a nonlinear optical process. The result is shown in Figure \ref{fig:dm_example}. The reconstruction with the tQST protocol required only 6 measurements instead of 16, achieving a fidelity with the noiseless target state of $F = 0.93$.

We then applied the same models to the purity estimation, and evaluated their efficacy by means of two figures of merit. The first one is the mean squared error (MSE) on the test set, defined as:
\begin{align}
    \text{MSE} = \frac{1}{N_{test}} \sum_{i} \lvert y_{p,i} - y_{o,i} \rvert^2,
\end{align}
with $y_{p,i} \left( y_{o,i} \right)$ the predicted (observed) purity, and $N_{test}$ the number of samples of the test set as above. The second one is the $R^2$ coefficient, defined as:
\begin{align}
    R^2 = 1 - \frac{\sum_i \left( y_{o,i} - y_{p,i} \right)^2}{\sum_i \left( y_{o,i} - \overline{y}_o \right)},
\end{align}
with $\overline{y}_o$ the average observed purity computed on the test set. The MSE of the \textcolor{black}{MLP} and \textcolor{black}{PEMLP} were $0.0144$ and $0.0097$, respectively. The corresponding $R^2$ coefficients were 0.7808 and 0.8528. 

These results demonstrate that both the \textcolor{black}{MLP} and the \textcolor{black}{PEMLP} can estimate the state purity from a limited number of measurements selected according to the tQST protocol. This is significant because, in principle, one would first need to reconstruct the density matrix via quantum state tomography and then compute the purity. If we reconstruct the density matrices corresponding to the test set with maximum likelihood and then estimate the purity, we find MSE = 0.0006 and $R^2$ = 0.9902.

We studied the performance of the models under the effect of the same noisy channels as in tQST. The results, reported in Appendix \ref{app:2qubit_results}, confirmed that both models are robust, with their performance being only slightly affected by increasing the noise level.

It is interesting to note that the \textcolor{black}{MLP} works better in tQST, while the \textcolor{black}{PEMLP} gives better results in the purity estimation. Maximum likelihood provides more accurate results when applied to both tasks. Nevertheless, the results on the 2-qubit datasets are a good benchmark that assess the capability of the models to characterize quantum states. In the next section we study the scalability of our approach to 4 qubits.

\subsection{4 Qubits}

Based on the results on the 2-qubit datasets, we decided to combine the two models to enhance the performance and conducted an ablation study by varying the number of hidden layers and the number of neurons/internal features per layer for each model separately\cite{cohen1988evaluation}. The results presented in Table \ref{tab:ablation_study} of Appendix \ref{app:ablation_study} clearly indicate that neither model, in isolation, can extract the relevant information from the input data to reliably reconstruct the density matrix.

\begin{figure*}
    \includegraphics[scale=0.3]{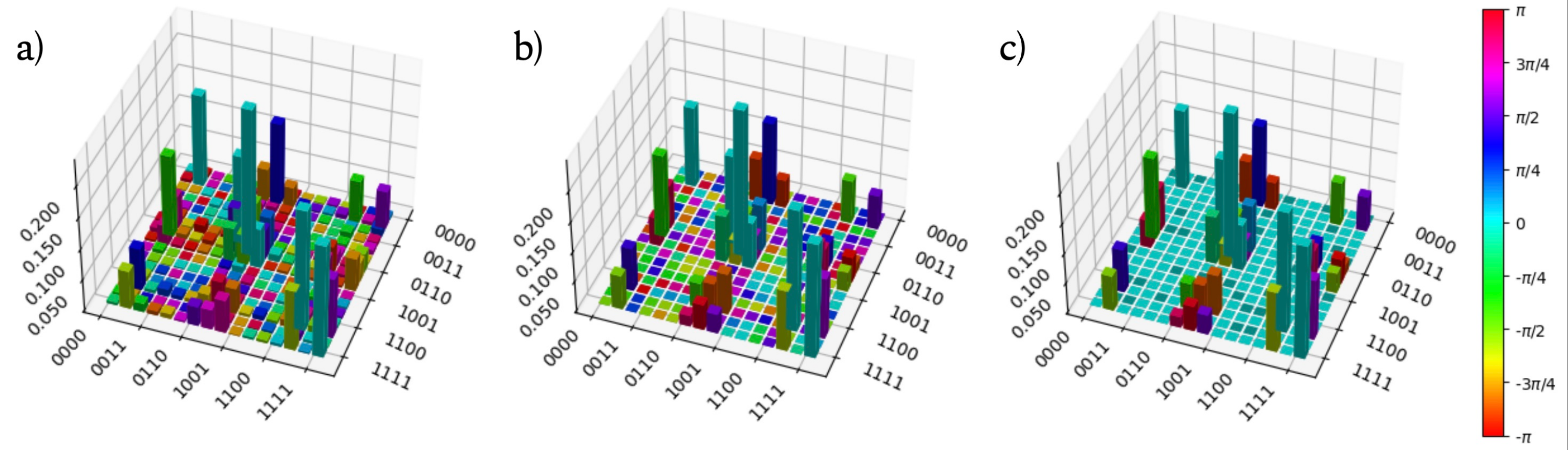}
    \caption{Reconstruction of a 4-qubit density matrix. The encoding of modulus and phase is the same as Figure \ref{fig:dm_example}. a) tQST reconstruction with the combined model trained on the noiseless dataset. The fidelity with the noiseless target state is 0.95. b) tQST reconstruction using maximum likelihood. The fidelity with the target density matrix is 0.99. c) Target density matrix.}
    \label{fig:dm_4qbit}
\end{figure*}

For tQST, the model showing the best trade-off between number of parameters, training time, and performance is composed of three hidden PELinear layers, with 32 internal features each, followed by one hidden Linear layer with 512 neurons. The activation function between each pair of hidden layers is the ReLU function. As in the previous case, the model is evaluated by computing the average fidelity and standard deviation on the test set. The average fidelity of the combined model on the noiseless test set is $0.8811 \pm 0.0540$. The result shows that the combined model can perform threshold quantum state tomography and reconstruct the density matrix of 4-qubit states, contrary to individual models. Maximum likelihood achieved an average fidelity of $F = 0.9982 \pm 0.0025$ on the noiseless test set. In Figure \ref{fig:dm_4qbit} we show a density matrix reconstructed with our model along with that obtained with the maximum likelihood and the target state. We performed only 46 measurements instead of 256 to reconstruct this density matrix.

We believe that the success of the combined model can be explained as follows. Thanks to the equivariance to permutations, the PELinear layers exploit our prior knowledge about the symmetric structure of density matrices. Unlike general linear layers, PELinear layers do not need to learn the relationship among the matrices' elements for each possible permutation. In practice, for $n$ qubits, PELinear layers are $(2^n)!$ times more sample-efficient than general Linear layers. This fact is reflected in the reduction of the number of learnable parameters (which does not depend on $n$) and in a better generalization accuracy. The PELinear layers essentially elaborate ``local information'', as the extracted feature vectors are ``located'' in the position occupied by the elements they refer to. The Linear layers combine all the features holistically, providing a more reliable final prediction.

The sparsity of the density matrix plays a crucial role in the tQST protocol. On the one hand, the larger the number of vanishing diagonal elements, the smaller the number of measurements we have to perform according to the tQST protocol. On the other hand, many of the states used in quantum communication or computation protocols are sparse in the computational basis. Thus, as we have done for the 2-qubit case, we analyzed the performance of the model as a function of the number of zeros on the diagonal of the density matrix. We separately computed the average fidelity for states with a small (0-4), intermediate (5-9), and large (10-14) number of zeros on the diagonal. The average fidelity was 0.88, 0.89, and 0.94, respectively. These results show that the combined model better reconstructs states that are more sparse and agrees with more general results on the tQST protocol\cite{binosi2024tailor}.

For the purity estimation, the model is the same as for tQST, with the addition of a dropout layer before the final Linear layer with probability equal to 0.5. We included this to avoid overfitting that came to light in initial numerical experiments.

On the noiseless dataset, the combined model achieved MSE = $0.0023$ and $R^2 = 0.9837$. These results demonstrate that the model can effectively estimate the purity of a quantum state using only the measurement outcomes required by the tQST protocol. Furthermore, they confirm that the selected measurements provide sufficient information to accurately estimate a quantity dependent on the density matrix. If we leverage maximum likelihood to reconstruct the density matrices and estimate the state purity, we achieve MSE = $2.4 \times 10^{-5}$ and $R^2 = 0.9998$.

\begin{figure}
    \includegraphics[scale=0.45]{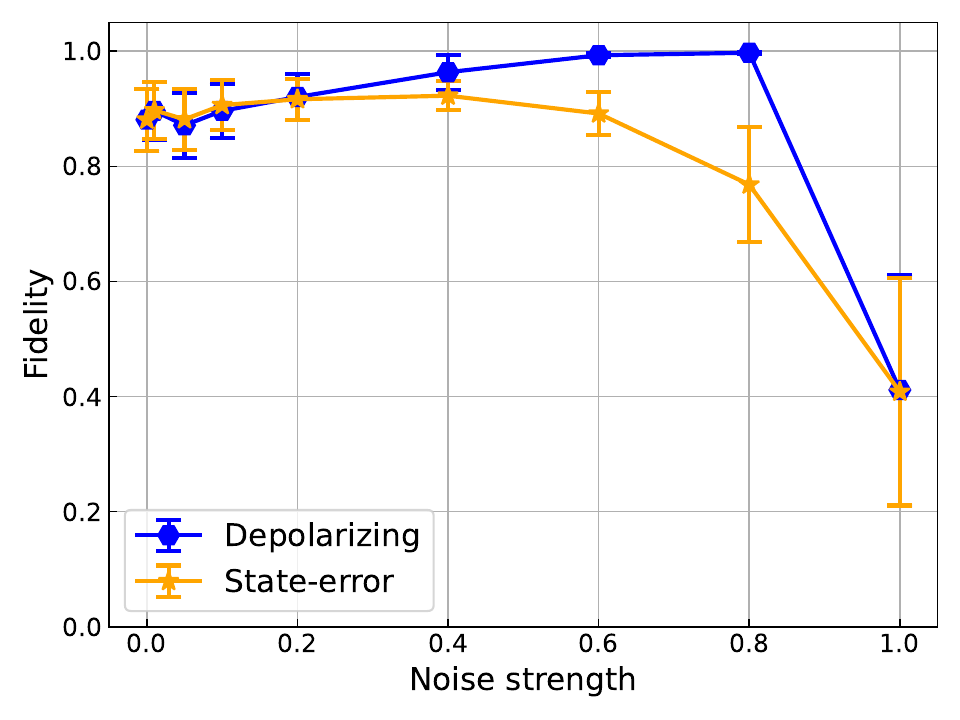}
    \caption{Fidelity on the test set with respect to noise strength for 4 qubits. The error bars are 1 standard deviation from the average.}
    \label{fig:fid_noise}
\end{figure}

\begin{figure}
    \includegraphics[scale=0.45]{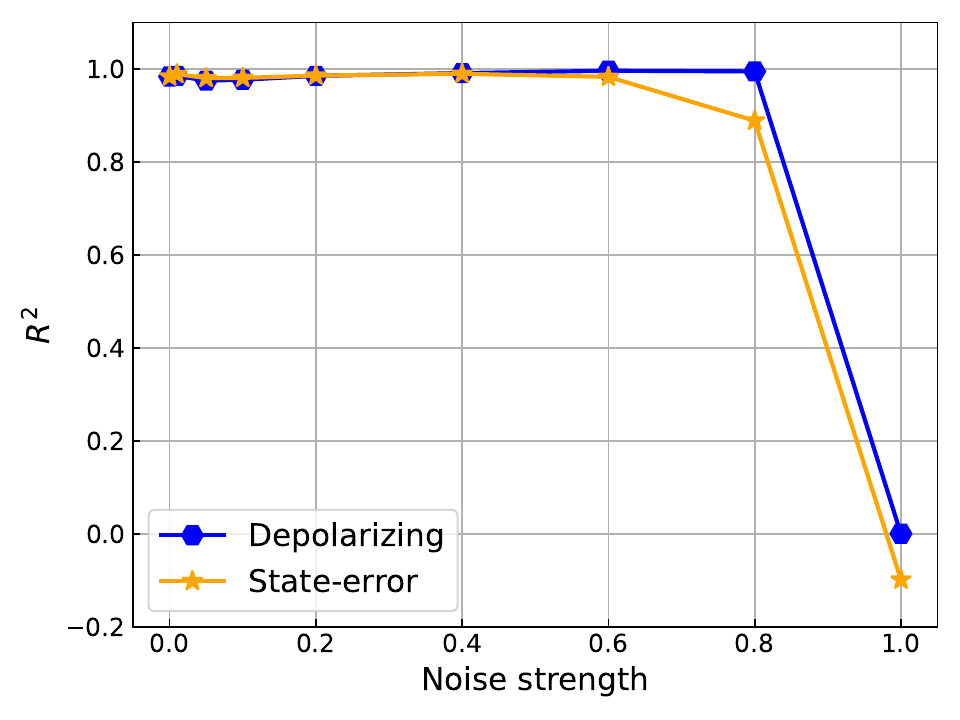}
    \caption{$R^2$ coefficient on the test set with respect to noise strength for 4 qubits.}
    \label{fig:rsquared_noise}
\end{figure}

We finally studied the impact of noise on the model performance. In our simulation the strengths $p$ and $\varepsilon$ of the noisy channels were set to 0.01, 0.05, 0.1, 0.2, 0.4, 0.6, 0.8, and 1.0. Figures \ref{fig:fid_noise} and \ref{fig:rsquared_noise} illustrate the performance of the combined model when trained on noisy data to reconstruct the noiseless density matrix or state purity, depending on the task. We can identify four regions, covering different ranges of noise strength. In the first region, from 0 to 0.1, the effect of the noise is small and comparable with statistical fluctuations. In the second one, ranging from 0.1 to 0.4, the noise provides a clear improvement of the results with respect to the noiseless case. \textcolor{black}{Although counterintuitive, this is a well-documented phenomenon in machine learning. Adding controlled noise can act as a form of regularization, preventing overfitting and improving generalization by encouraging the model to learn more robust features\cite{sietsma1991creating, bishop1995training, an1996effects, goodfellow2016deep, benedetti2024training}.} The performance of the model in the first two regions is similar for either the depolarizing or the state-error noise channel. On the contrary, in the third region, between 0.4 and 0.8, the model works much better in the case of a depolarizing channel. When the noise increase, the number of measurements approaches 256 for basically every test sample. However, the model can counteract the depolarizing noise more effectively because of its deterministic nature. In other words, unlike the case of state-error noise, which is random, the model is able to learn the correlations that characterize the depolarizing noise. Finally, in the last region, where $p=1$ and $\varepsilon=1$, the measurement outcomes do not provide sufficient information to successfully complete the tasks.


\section{Conclusions}

In this work, we leveraged the tQST protocol and the permutation-equivariance property to develop a deep learning model for quantum state characterization. We tested our model on tQST and purity estimation tasks, showing results for two- and four-qubit datasets. Our model successfully reconstructed density matrices and estimated purities, demonstrating robustness against depolarizing and state-error imperfections. \textcolor{black}{Our proposed model can compete with MLE, achieving compatible results in many cases. A key advantage emerges when noise is present. Our model can reconstruct the noiseless state from noisy measurement outcomes, whereas MLE can only retrieve the noisy state. As a result, MLE's performance degrades as noise increases, while our model exhibits robustness in tQST, as explained in Sec. \ref{sec:results}. For example, if we consider depolarizing noise with $p=0.4$, our model achieves an average fidelity of $0.96 \pm 0.03$, while MLE obtains $0.83 \pm 0.05$. Furthermore, the inference times of our model after training and MLE are comparable, about a few seconds.} 
\textcolor{black}{We also extended the applicability of the tQST protocol to estimate the state purity from the measurement outcomes, without reconstructing the density matrix. This is a valuable result of our work for at least two reasons. First, we showed that the tQST protocol provides meaningful measurements not only to reconstruct the density matrix, but also to estimate the state purity. Second, having an approach to directly estimate the purity can avoid the computational cost of MLE to reconstruct the density matrix for a large number of qubits.}

\textcolor{black}{In our study, we show the feasibility of embedding permutation-equivariant deep learning models into the threshold quantum state tomography protocol, providing a promising method for quantum state characterization in realistic scenarios. Our work also demonstrates the ability of machine learning to manage noisy measurement results in quantum state characterization, ensuring both robustness and accuracy despite experimental imperfections. While our machine learning approach does not outperform MLE, we believe our results are sufficiently encouraging to suggest more research in this direction.}
The results suggest that the integration of physics-inspired architectures and advanced computational techniques could substantially reduce resource requirements and improve the fidelity of quantum state reconstructions. This advancement not only provides a valuable tool for current quantum technology applications but also lays the groundwork for further exploration into larger and more complex quantum systems. Future research is essential to refine these models, explore optimal thresholds, \textcolor{black}{find more efficient ways to handle input data,} and extend their applicability to \textcolor{black}{many}-qubit systems, ensuring readiness for the demands of practical quantum computation and communication technologies. Yet, this is beyond the scope of this work.

Still, our promising results on noisy data highlight the potential of a machine learning approach for quantum state characterization, especially in the context of tQST, which enables significant reductions in the number of required measurements. These findings underline the importance of further studies to address the scalability of this approach for very large quantum systems.


\medskip
\noindent{\bf Funding:} This work is supported by PNRR MUR project PE0000023-NQSTI.

\section*{Author Declarations}

\subsection*{Conflict of Interest}
The authors have no conflicts to disclose.

\subsection*{Author Contributions}

\textbf{D. Maragnano}: Conceptualization (equal); Data curation (equal); Formal analysis (equal); Investigation (equal); Methodology (equal); Software (Equal); Writing - original draft (equal); Writing - review and editing (equal). \textbf{M. Liscidini}: Conceptualization (equal); Formal analysis (equal); Investigation (equal); Methodology (equal); Funding acquisition (lead); Investigation (equal); Methodology (equal); Project administration (lead); Writing - original draft (equal); Writing - reviewing and editing (equal). \textbf{C. Cusano}: Conceptualization (supporting); Data curation (equal); Formal analysis (equal); Investigation (equal); Methodology (equal); Software (Equal); Writing - original draft (equal); Writing - reviewing and editing (equal).

\section*{Data Availability}

\textcolor{black}{The data that support the findings of this study are openly available on Zenodo at \href{https://doi.org/10.5281/zenodo.14887723}{https://doi.org/10.5281/zenodo.14887723}.}

\appendix

\section{Neural Networks for tQST}\label{app:nns}

Artificial Neural Networks are universal approximators~\cite{hornik1989multilayer}, that is, they can replicate arbitrarily well any target behavior. However, just relying on this property may impose a very high cost in terms of the complexity of the model, computational burden of the training procedure, and number of training samples. To control these factors, we can imbue the architecture with prior knowledge of the mathematical structure of the optimal solution.

To this purpose, many recent studies explored the concept of equivariance in neural networks~\cite{ravanbakhsh2017equivariance,kondor2018generalization,yarotsky2021universal}. Briefly, a network (or one of its components) is equivariant to a group of transformations if applying a transformation in the group to the input causes a predictable change in the output. For instance, convolutional neural networks (CNN) are equivariant to translations in the plane, and graph neural networks (GNN) are equivariant to permutations in the order used to label the vertices in the graphs~\cite{maron2018invariant}.

Here, we consider the case of neural networks applied to density matrices.
For density matrices, the permutation of rows and columns should result in a similarly permuted output from the neural network. This recalls the case of adjacency matrices processed by GNNs.

In this context, a neural network $F: \mathbb{C}^{n \times n} \to \mathbb{C}^{n \times n}$ is \emph{permutation equivariant} if for any input matrix $\rho$, and any permutation matrix $\pi$, we have:
\begin{equation}
    \label{eq:equivariance}
    F(\pi^T \rho \pi) = \pi^T F(\rho) \pi.
\end{equation}
A network of this kind can be built as a sequence of permutation equivariant layers, each one enjoying property (\ref{eq:equivariance}). The general form for permutation equivariant linear layers~\cite{thiede2020general} has been derived by Thiede \emph{et al.} They showed that any permutation equivariant linear operator $\rho \to \rho'$  can be expressed as follows:
%
%
\begin{equation}
    \begin{split}
        \rho'_{ij} &=
        w_0 \rho_{ij} + 
        w_1 \rho_{ji} + 
        w_2 \sum_v \rho_{iv} + 
        w_3 \sum_v \rho_{vi} + 
        w_4 \sum_u \rho_{uj} \\
        &+ w_5 \sum_u \rho_{ju} + 
        w_6 \sum_{u,v} \rho_{uv} + 
        w_7 \sum_u \rho_{uu} + 
        w_8 \rho_{ii} + 
        w_9 \rho_{jj} \\ &+
        \delta_{ij} \left(
        w_{10} \rho_{ij} +
        w_{11} \sum_u \rho_{uu} +
        w_{12} \sum_{u,v} \rho_{uv}\right. \\ &+ 
        \left. w_{13} \sum_v \rho_{iv} + 
        w_{14} \sum_v \rho_{vi}
        \right),
    \end{split}
\end{equation}
where $w_0, w_1, \dots, w_{14}$ are the learnable parameters of the operator, and $\delta_{ij} = 1$ if $i = j$, $\delta_{ij} = 0$ if $i \neq j$.

In practice, this formulation is extended to the case where matrix elements are vectors of real numbers: $\rho_{ij} \in \mathbb{R}^c$, $\rho_{ij}' \in \mathbb{R}^d$, and all coefficients are matrices of real-valued learnable parameters: $w_k \in \mathbb{R}^{c \times d}$.

To build a neural network, we need more than just linear layers. Luckily, permutation equivariance is closed under functional composition, and most auxiliary layers in neural networks are already permutation equivariant (activation functions, normalization, and pooling, for instance). This means that a complete permutation-equivariant network can be obtained as a sequence of permutation-equivariant linear layers and other auxiliary layers.

The networks we experimented with receive complex numbers as input. These are encoded as two-dimensional vectors before processing.

\section{Training Details}\label{app:training_details}

The code for the training of the models was written in Python using the Pytorch library \cite{paszke2019pytorchimperativestylehighperformance}.
All the datasets were split into training, validation, and test sets with shares equal to 90\%, 5\%, and 5\%, respectively, and loaded with batch size equal to 32.

Both the threshold quantum state tomography and the purity estimation are regression task. Thus, we adopted the MSELoss as objective function to be minimized. We used the Adam optimizer with learning rate equal to $10^{-4}$. The number of epochs was 100.

The 2-qubit \textcolor{black}{MLP} models for the tomography and purity estimation have 2128 and 1633 parameters, respectively, and required a training time of $\simeq$ 2 minutes on the Google Colab platform without GPUs. The \textcolor{black}{PEMLP} models have 3972 and 3989 parameters for the same tasks, and required $\simeq$ 30 minutes for the training on the same platform. 

The 4-qubit \textcolor{black}{combined} model for tQST has 295,748 parameters, while the one for the purity has 164,933 parameters, and both required $\simeq$ 4 hours for the training on a standard desktop computer with an Intel Core i7-4790 processor, 3.60 GHz CPU, and 32 GB system memory, without GPUs.

\section{2-qubit results}\label{app:2qubit_results}
In this appendix we report the complete results for the 2-qubit tasks. In Table \ref{tab:2q_tqst} and Table \ref{tab:2q_purity} we show the results for the \textcolor{black}{MLP} and \textcolor{black}{PEMLP} on the tQST and purity estimation, respectively.

\begin{table*}
\caption{\label{tab:2q_tqst} Results for the \textcolor{black}{MLP} and \textcolor{black}{PEMLP} on the 2-qubit noiseless and noisy datasets on the tQST. For each value of the noise strength we reported the average fidelity and the standard deviation on the test set.}
\begin{ruledtabular}
\begin{tabular}{cccc}
& \textcolor{black}{MLP} - tQST & \\
\hline
Noise strength & Depolarizing & State-error \\
\hline
0    & 0.9555 / 0.0411 & 0.9555 / 0.0411 \\
0.01 & 0.9501 / 0.0505 & 0.9624 / 0.0353 \\
0.05 & 0.9561 / 0.0405 & 0.9448 / 0.4978 \\
0.1  & 0.9578 / 0.0436 & 0.9366 / 0.0591 \\
\hline
& \textcolor{black}{PEMLP} - tQST & \\
\hline
Noise strength & Depolarizing & State-error \\
\hline
0    & 0.8893 / 0.0919 & 0.8893 / 0.0919 \\
0.01 & 0.8918 / 0.0907 & 0.8838 / 0.0916 \\
0.05 & 0.8864 / 0.0915 & 0.8901 / 0.0907 \\
0.1  & 0.8883 / 0.0900 & 0.8858 / 0.9120 \\
\end{tabular}
\end{ruledtabular}
\end{table*}

\begin{table*}
\caption{\label{tab:2q_purity} Results for the \textcolor{black}{MLP} and \textcolor{black}{PEMLP} on the 2-qubit noiseless and noisy datasets on the purity estimation. For each value of the noise strength we reported the MSE and the $R^2$ on the test set.}
\begin{ruledtabular}
\begin{tabular}{cccc}
& \textcolor{black}{MLP} - Purity estimation & \\
\hline
Noise strength & Depolarizing & State-error \\
\hline
0    & 0.0144 / 0.7808 & 0.0144 / 0.7808 \\
0.01 & 0.0163 / 0.7523 & 0.0148 / 0.7739 \\
0.05 & 0.0157 / 0.7615 & 0.0160 / 0.7562 \\
0.1  & 0.0162 / 0.7536 & 0.0139 / 0.7884 \\
\hline
& \textcolor{black}{PEMLP} - Purity estimation & \\
\hline
Noise strength & Depolarizing & State-error \\
\hline
0    & 0.0097 / 0.8528 & 0.0097 / 0.8528 \\
0.01 & 0.0101 / 0.8462 & 0.0123 / 0.8123 \\
0.05 & 0.0104 / 0.8420 & 0.0118 / 0.8208 \\
0.1  & 0.0106 / 0.8382 & 0.0149 / 0.7734 \\
\end{tabular}
\end{ruledtabular}
\end{table*}

\section{Ablation Study}\label{app:ablation_study}

In this appendix we report the complete results of the ablation study performed on the 4-qubit dataset. We considered different numbers of hidden layers and neurons/ internal features per hidden layer for both models. The activation function between the layers was the ReLU function. We trained each model on the noiseless dataset. The results are reported in Table \ref{tab:ablation_study}.

\begin{table*}
\caption{\label{tab:ablation_study}Ablation study performed on the \textcolor{black}{MLP} and \textcolor{black}{PEMLP} on the 4-qubit dataset. For each configuration we reported the average fidelity and the standard deviation on the test set.}
\begin{ruledtabular}
\begin{tabular}{cccccc}
 & & \textcolor{black}{MLP} & & &  \\
\hline
Number of neurons per hidden layer  & 64 & 128 & 256 & 512 & 1024\\
\hline
1 hidden layer & 0.4282 / 0.1951 & 0.5094 / 0.1829 & 0.5926 / 0.1842 & 0.6098 / 0.1840 & 0.6745 / 0.1505 \\
2 hidden layers & 0.6072 / 0.1861 & 0.6093 / 0.1853 & 0.6109 / 0.1842 & 0.6414 / 0.1672 & 0.6424 / 0.1668 \\
3 hidden layers & 0.6185 / 0.1783 & 0.6441 / 0.1638 & 0.6775 / 0.1645 & 0.7102 / 0.1254 & 0.6585 / 0.1586 \\
\hline
 & & \textcolor{black}{PEMLP} & & & \\
\hline
Number of internal features per hidden layer  & 8 & 16 & 32 & 64 & 128 \\
\hline
1 hidden layer & 0.6105 / 0.1857 & 0.6132 / 0.1846 & 0.6067 / 0.1846 & 0.6164 / 0.1837 & 0.6207 / 0.1812  \\
2 hidden layers & 0.6064 / 0.1855 & 0.6043 / 0.1845 & 0.6033 / 0.1837 & 0.5985 / 0.1800 & 0.5985 / 0.1766 \\
3 hidden layers & 0.4114 / 0.1994 & 0.6169 / 0.1807 & 0.6011 / 0.1819 & 0.6239 / 0.1766 & 0.6338 / 0.1726 \\
\end{tabular}
\end{ruledtabular}
\end{table*}

\nocite{*}
\bibliography{main}

\end{document}